\documentclass[twocolumn]{aastex62}
\newcommand{\psr}{PSR~J1431$-$6328}
\newcommand{\askap}{ASKAP~143121.2$-$632809}

\shorttitle{Serendipitous Discovery of PSR~J1431$-$6328}
\shortauthors{Kaplan et al.}

\submitjournal{ApJ}
\accepted{August 6, 2019}

\begin{document}

\title{Serendipitous Discovery of PSR~J1431$-$6328 as a Highly-Polarized Point Source with the Australian SKA Pathfinder}

\author[0000-0001-6295-2881]{David~L.~Kaplan}
\affiliation{Center for Gravitation, Cosmology, and Astrophysics, Department of Physics, University of Wisconsin-Milwaukee, P.O. Box 413, Milwaukee, WI 53201, USA}

\author[0000-0002-9618-2499]{Shi Dai}
\affiliation{ATNF, CSIRO Astronomy and Space Science, PO Box 76, Epping, New South Wales 1710, Australia}

\author[0000-0002-9994-1593]{Emil Lenc}
\affiliation{ATNF, CSIRO Astronomy and Space Science, PO Box 76, Epping, New South Wales 1710, Australia}

\author[0000-0002-9583-2947]{Andrew Zic}
\affiliation{Sydney Institute for Astronomy, School of Physics, University of Sydney, Sydney, New South Wales 2006, Australia.}
\affiliation{ATNF, CSIRO Astronomy and Space Science, PO Box 76, Epping, New South Wales 1710, Australia}

\author[0000-0002-1075-3837]{Joseph~K.~Swiggum}
\affiliation{Center for Gravitation, Cosmology, and Astrophysics, Department of Physics, University of Wisconsin-Milwaukee, P.O. Box 413, Milwaukee, WI 53201, USA}

\author[0000-0002-2686-438X]{Tara Murphy}
\affiliation{Sydney Institute for Astronomy, School of Physics, University of Sydney, Sydney, New South Wales 2006, Australia.}

\author[0000-0002-6243-7879]{Craig~S.\ Anderson}
\affiliation{CSIRO Astronomy \& Space Science, 26 Dick Perry Avenue, Kensington, WA 6151, Australia}

\author[0000-0002-2037-4216]{Andrew D.~Cameron}
\affiliation{ATNF, CSIRO Astronomy and Space Science, PO Box 76, Epping, New South Wales 1710, Australia}

\author[0000-0003-0699-7019]{Dougal Dobie}
\affiliation{Sydney Institute for Astronomy, School of Physics, University of Sydney, Sydney, New South Wales 2006, Australia.}
\affiliation{ATNF, CSIRO Astronomy and Space Science, PO Box 76, Epping, New South Wales 1710, Australia}

\author[0000-0003-1502-100X]{George Hobbs}
\affiliation{ATNF, CSIRO Astronomy and Space Science, PO Box 76, Epping, New South Wales 1710, Australia}

\author[0000-0003-4810-7803]{Jane F.~Kaczmarek}
\affiliation{ATNF, CSIRO Astronomy and Space Science, PO Box 76, Epping, New South Wales 1710, Australia}
\affiliation{CSIRO Astronomy and Space Science, Parkes Observatory, P.O. Box 276, Parkes NSW 2870, Australia}

\author[0000-0002-0494-192X]{Christene Lynch}
\affiliation{International Centre for Radio Astronomy Research - Curtin University, 1 Turner Avenue, Bentley, WA 6102, Australia}
\affiliation{ARC Centre of Excellence for All Sky Astrophysics in 3 Dimensions (ASTRO 3D)}

\author[0000-0003-3186-3266]{Lawrence Toomey}
\affiliation{ATNF, CSIRO Astronomy and Space Science, PO Box 76, Epping, New South Wales 1710, Australia}

\begin{abstract}
We  identified a highly-polarized, steep-spectrum radio source in a
deep image with the Australian Square Kilometre Array Pathfinder
(ASKAP) telescope at 888\,MHz.  After considering and rejecting a
stellar origin for this source, we discovered a new millisecond pulsar (MSP)
using observations from the Parkes radio telescope.  This pulsar has
period 2.77\,ms and dispersion measure $228.27\,{\rm pc\,cm}^{-3}$.
Although this pulsar does not yet appear to be particularly remarkable, the short spin period, wide profile and high dispersion measure do make it relatively hard to discover through traditional blind periodicity searches.
Over the course of several weeks we see changes in the barycentric period of this pulsar that are consistent with orbital motion in a binary system, but the properties of any binary need to be confirmed by further observations. While even a deep ASKAP survey may not identify large numbers of new MSPs compared to the existing population, it would be competitive with existing all-sky surveys and could discover interesting new MSPs at high Galactic latitude without the need for computationally-expensive all-sky periodicity searches.
\end{abstract}


\section{Introduction}

While traditional pulsar searches employ time-domain techniques to
identify periodic signals \citep[e.g.,][]{2012hpa..book.....L}, these
techniques are limited: eclipses, scattering, or orbital motion can
all negatively impact  time-domain searches  even when the underlying
radio emission remains constant.  Separately, even when pulsed
emission is detectable, non-traditional searches can help explore
large regions of parameter space efficiently, and imaging searches in general can focus resources on the sources of greatest interest.
Identifying steep-spectrum radio sources has long been used to identify new pulsars, and in fact has been able to help identify new classes of pulsars before they were found in more traditional ways \citep[][and references therein]{1979ApJ...228..755R,1980BAAS...12..799E,1995ApJ...455L..55N,1987ApJ...319L.103S,2016MNRAS.461.1062F}.
This technique has found new life recently
with  both targeted observations \citep[e.g.,][]{2019ApJ...876...20H} and more importantly large-scale radio surveys  \citep[as in][]{2018MNRAS.475..942F}, in some cases guided by other ``pulsar-like" objects of interest identified through multi-wavelength surveys.

Beyond steep-spectrum catalogs, new techniques are also under consideration. For instance, \citet{2016MNRAS.462.3115D} discuss identifying pulsar-like sources through their interstellar scintillation properties.  Here we discuss another way to identify pulsars from large-scale radio surveys: by searching for the circularly polarized emission that is almost unique to pulsars (as mentioned in \citealt{2018MNRAS.475..942F} and others), although even among pulsars it is not universal \citep{1998MNRAS.300..373H,1998MNRAS.301..235G}.
A small number of pulsars have been detected through a combination of spectral properties and linear polarization \citep{1995ApJ...455L..55N,1987ApJ...319L.103S}, but there have not been large-scale surveys including circular polarization until recently \citep{2018MNRAS.478.2835L}.
Very few radio sources have more than a few percent circular polarization at frequencies $<5\,$GHz.  Many types of stars show some radio emission
\citep{2002ARA&A..40..217G}, although for most of them it is weak,
incoherent emission.  Low-mass flare stars 
typically show bursty emission that is often highly polarized \citep[e.g.,][]{2019ApJ...871..214V,zic19}.  Likewise, chromospherically-active binaries like RS~CVns show polarized flares \citep[e.g.,][]{1987AJ.....93.1220M}.
While some steady emission has been seen from these sources 
\citep{1994ApJS...90..743G},
the majority of polarized sources seen in large-scale surveys are pulsars
\citep[e.g.,][]{2018MNRAS.478.2835L}.

Here we report the serendipitous discovery of a highly-polarized steep spectrum point source in a deep pointing with the Australian Square Kilometre Array Pathfinder telescope
 \citep[ASKAP;][]{2014PASA...31...41H,2016PASA...33...42M}
 at 888\,MHz.  After considering and rejecting a flare-star origin we discovered a new pulsar in a dedicated pointing with the Parkes radio telescope.  We discuss the nature of this pulsar and prospects for identifying future sources through polarization imaging surveys.

\section{ASKAP Observations and Source Discovery}
\begin{figure*}
\epsscale{1.2}
  \plotone{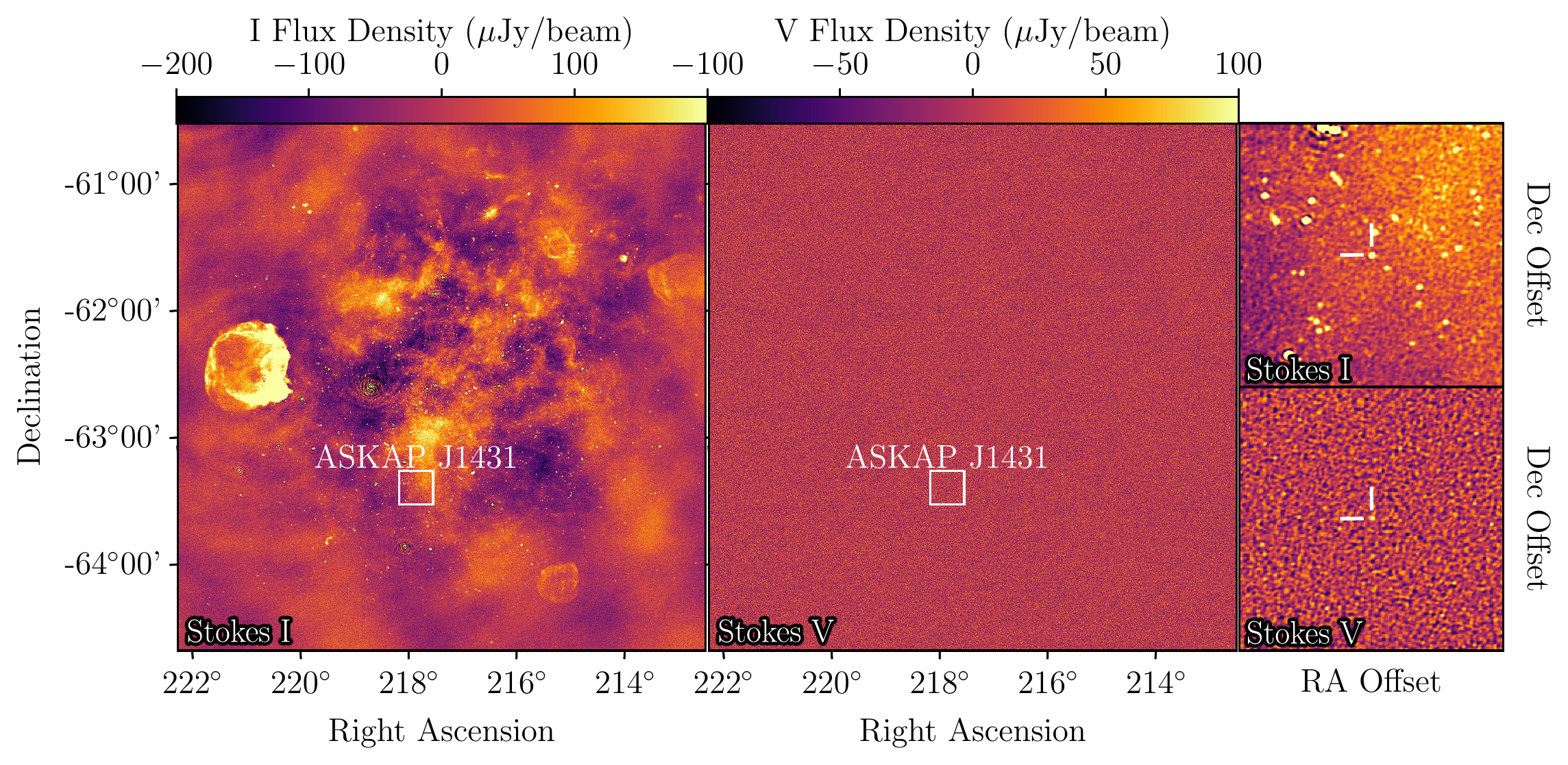}
  \caption{ASKAP 888\,MHz image of the field centered on Proxima
    Centauri, from 2019~May~1.  We show the total intensity (Stokes I) image on the
    left, and the circular polarization (Stokes V) image in the
    center.  Each is $250\arcmin$ on a side, with north up and east to
    the left.  The regions of the zoomed images highlighting \askap\ (labeled as ``ASKAP~J1431") are
    shown with boxes.  On the right we show the zoomed image in Stokes
    I (top) and Stokes V (bottom).  Each image is $8\arcmin$  on a side, with north up and east to
    the left.  The color scales are the same as the larger Stokes I
    and Stokes V images.  We indicate the position of \askap\ with tick
    marks. Note that these images have not been corrected for primary
    beam attenuation. }
  \label{fig:askap}
\end{figure*}

As part of a campaign to study the nearby flare star \object[V* V645
  Cen]{Proxima Centauri} (Zic et al., in prep.), we observed a field centered at (J2000)
$\alpha=14^{\rm h}29^{\rm m}32\fs32$,
$\delta=-62\degr40\arcmin31\farcs3$ with ASKAP on each night from 2019~May~1
through 2019~May~4, as detailed in Table~\ref{tab:obs}.  The observations used an integration time of 10\,s with baselines ranging from 22.4\,m to 6.4\,km.  We reduced the data following the procedure in \citet{zic19}, using the Common Astronomy Software Applications package \citep[\textsc{casa},][]{casa} to make images in circular polarisation (Stokes V) as well as total
intensity (Stokes I). The primary calibrator PKS B1934$-$638 was observed for each observing epoch to calibrate the flux scale and the instrumental bandpass.  The resulting images have synthesized beam widths of $12\arcsec-14\arcsec$, and were created with a robustness of 0.  The images from the first observation are shown
in Figure~\ref{fig:askap}.  The rms noise
in the  images is around $47\,\mu$Jy/beam (Stokes V) and
$57\,\mu$Jy/beam (Stokes I).

\begin{figure}
  \plotone{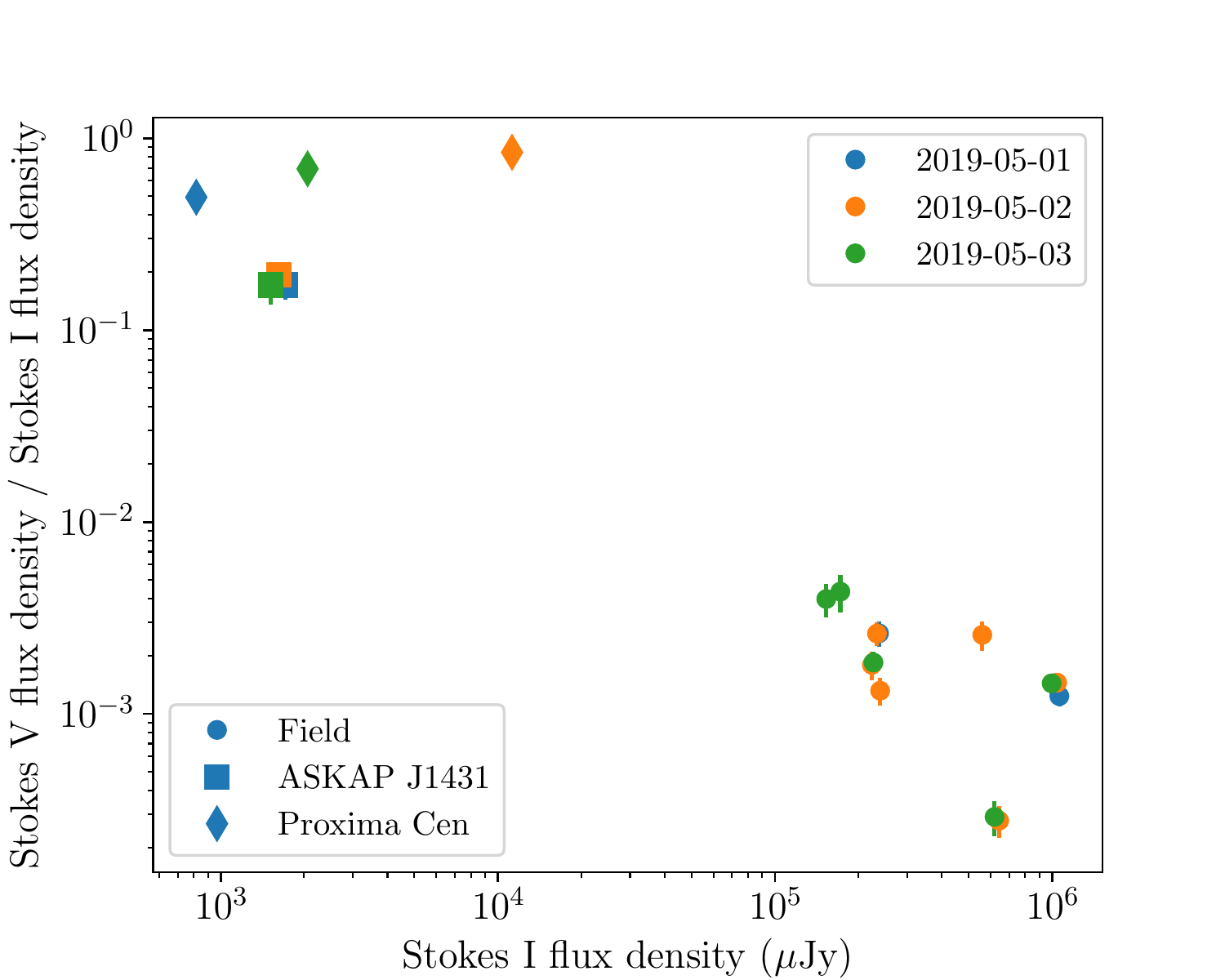}
    \caption{Fractional circular polarization  in our ASKAP images.  We
      show the $V/I$ flux density fraction against primary-beam
      corrected Stokes I flux
      density for the first three observations.  All sources are detected at $>5\sigma$ significance in the Stokes V images.  Proxima Centauri is
      plotted as a diamond, \askap\ as a square, and the remaining field
      sources (dominated by leakage) as circles.  Observation date is indicated by color (2019~May~1 is blue, 2019~May~2 is orange, 2019~May~3 is green).}
    \label{fig:leakage}
\end{figure}

The total intensity image is quite complex, with over 3000 point
sources visible along with significant diffuse emission and several
supernova remnants (the brightest of which is \object{RCW 86}).
However, the circular polarization image has fewer than 10 sources
visible.   To quantify this, we identified all of the significant
sources in the Stokes V images using \texttt{aegean}
\citep{2018PASA...35...11H}, and matched them
against sources in the Stokes I images.   We show the results in
Figure~\ref{fig:leakage}.  With the exception of two sources, all of
the Stokes V detections are bright sources (flux
densities $>10^5\,\mu$Jy) with Stokes V detectable because of a modest
level of leakage, typically 0.1\%.  The two exceptions are marked
separately.  One is Proxima~Cen, which was known to be polarized.  The
other was a previously unknown radio source which we designate \askap.

This source has Stokes I flux density $1666\pm40\,\mu$Jy and Stokes V flux
density $304\pm34\,\mu$Jy, and does not
appear to vary between the images.  The large polarized fraction of
$V/I=18\pm2$\% is notable, and led us to investigate this source further.

To determine the spectral properties of this source we combined the
four observations into a single UV data set and then imaged it in four
sub-bands spread across the 288\,MHz bandpass.  With this we measure a
spectral index $\alpha=-1.5\pm0.2$ across the 780--995\,MHz region
(with flux density $S_\nu \propto \nu^{\alpha}$).  Such a steep spectrum is consistent with the median spectral index of pulsars \citep[e.g.,][]{2013MNRAS.431.1352B} and steeper than 90\% of radio sources \citep{2017A&A...598A..78I}.  However, note that the spectral index \textit{within} the ASKAP observation was only apparent after combining multiple observations and would be difficult to use on its own for source selection, while the high degree of polarization was much more apparent.  Comparing to other surveys would be better for steep spectrum selection, but there are few appropriate radio surveys below $\delta\approx -40\degr$.

The initial image astrometry showed some small ($<1\arcsec$) discrepancies between the positions of radio sources and their cataloged values that were nonetheless larger than the statistical uncertainties.  We therefore shifted the positions to better match two sources from the The Australia Telescope 20 GHz \citep[AT20G;][]{2010MNRAS.402.2403M} catalog.  After this shift 
we find a position for \askap\ at epoch 2019.33 of $14^{\rm h}31^{\rm m}21\fs23$,
$-63\degr28\arcmin08\farcs6$, based on the Stokes I images.  The statistical errors are about
$\pm0\farcs2$ on each coordinate, but we include an additional
$\pm0\farcs3$ uncertainty in each coordinate that comes from the match
of Proxima Centauri (based on
\citealt{2018A&A...616A...1G}) with its position.  
There
is also an additional $2\arcsec$ shift between the position of the
Stokes I source and the Stokes V source, although both are
individually consistent with point sources.  This is being further investigated by the ASKAP commissioning team.

\begin{deluxetable*}{l c c c c c c c}
\tablewidth{0pt}
\tablecaption{Observation Summary\label{tab:obs}}
\tablehead{\colhead{Telescope} & \colhead{Mode} & \colhead{Start} & \colhead{Duration} & \colhead{Frequency Range} & \colhead{Epoch\tablenotemark{a}} & \colhead{$P$\tablenotemark{a}}\\
 & & (UT) & (h) & (MHz) & \colhead{(MJD TDB)} & \colhead{(ms)}}
\startdata
ASKAP & Imaging & 2019~May~1 09:00 & 14.0 & 744--1032 & \nodata\\
ASKAP & Imaging & 2019~May~2 09:00 & 14.0 & 744--1032 & \nodata\\
ASKAP & Imaging & 2019~May~3 09:00 & 14.0 & 744--1032 & \nodata\\
ASKAP & Imaging & 2019~May~4 12:12 & \phn4.0 & 744--1032 & \nodata\\
ATCA & Imaging & 2019~May~17 10:18 & \phn4.3 & 4500--6500 & \nodata\\
Parkes & Pulsar & \dataset[2019~May~19 08:12]{\doi{10.25919/5d02232da5691}} & \phn1.8 & 704--4032 & 58622.34684021 & 2.77229565(2)\phn\\
Parkes & Pulsar & \dataset[2019~May~27 07:26]{\doi{10.25919/5d0222fcaf307}} & \phn2.0 & 704--4032 & 58630.31483975 & 2.77229303(2)\phn\\
Parkes & Pulsar & \dataset[2019~May~27 09:42]{\doi{10.25919/5d040bea8de5d}} & \phn1.5 & 704--4032& 58630.40936468 & 2.77229306(2)\phn\\
Parkes & Pulsar & \dataset[2019~May~28 11:05]{\doi{10.25919/5d0bf50244af7}} & \phn0.5 & 704--4032 & 58631.46696174 & 2.77229354(12)\\
Parkes & Pulsar & \dataset[2019-June~09 07:20]{\doi{10.25919/5d0c630dc299a}} & \phn0.4 & 704--4032 & 58643.31064798 & 2.7723076(2)\phn\phn\\
\enddata
\tablecomments{Numbers in parentheses are the 1$\sigma$ errors in the last digit quoted.}
\tablenotetext{a}{Barycentric MJD and  spin-period.}
\end{deluxetable*}

\section{Multi-wavelength Observations and Followup}
 To further constrain the  nature of \askap\ we observed the source with the Australia Telescope Compact Array (ATCA) on 2019~May~19  with a 2\,GHz bandwidth centered at 5.5\,GHz (Table~\ref{tab:obs}, project code CX436). As a result of a pointing error, the source was offset in the primary beam and this affected sensitivity. The source was not detected in Stokes I, adjusting for the primary beam attenuation we obtain a $3\sigma$ upper limit of $51\,\mu{\rm Jy\,beam}^{-1}$.  Combining this with the ASKAP measurements suggests a slightly steeper spectral index of $<-1.9$, compared to the ASKAP-only value of $-1.5\pm0.2$.  This confirms that either the source is  steep-spectrum, or that it varies; although variability seems to be inconsistent with the multiple ASKAP detections.

As mentioned above, two classes of stars that
show coherent, highly polarized emission are magnetic low-mass flare
stars and RS~CVn binaries.  To search for any possible stellar
counterpart to \askap, we searched the available archives and found
deep near-infrared images from the VISTA Variables in the Via Lactea (VVV; \citealt{2010NewA...15..433M}) survey as the best available data.  
We see no sources detected within that 3$\sigma$ error circle down to $5\sigma$ limiting magnitudes $J>20.4$\,mag, $H>19.9$\,mag, and $K_s>19.0$\,mag.

Based on these limits, we can constrain any possible low-mass star/substellar object.  Based on the observed population of ultra-cool dwarfs \citep{2008AJ....136.1290R} we determine a lower limit on the distance of any counterpart of 250--1000\,pc (for late L to mid-M spectral types).  From these we computed upper limits on the radio flux density, assuming $L_{\rm radio}/L_{\rm bol}=10^{-7}$, which is appropriate for M dwarfs  \citep{2010ApJ...709..332B}.  This gives limits of $<0.5\,\mu$Jy, compared with our detection of $1.66\,$mJy.  Even for later L dwarfs the ratio might increase to $10^{-5}$, which would imply a radio flux density of $50\,\mu$Jy that is still a factor of 30 below our detection.

Similarly, we can limit any RS~CVn binary.  \citet{1987AJ.....93.1220M} measured flux densities of $\approx 30\,$mJy for \object{UX Ari} at 1\,GHz, which has $J=4.5$\,mag \citep{2002yCat.2237....0D}.  Scaling that to our measured flux density would imply $J\approx 8$\,mag, which is incompatible with our VVV upper limit.
Even more so than with ultra-cool dwarfs, we can easily rule out any chromospherically active binary like RS~CVns.

Since this source seemed to not be stellar, we  observed the position of \askap\ on 2019~May~19 at Parkes using the recently-commissioned Ultra-Wideband Low receiver (UWL; \citealt{UWL}) on the 64-m Parkes radio telescope, which provides radio frequency coverage from 704\,MHz to 4032\,MHz.
The observation was performed using the transient search mode and the DFB4 backend, where data is recorded with 2-bit sampling every 64\,$\mu$s in each of the 0.5\,MHz channels covering 1241--1497\,MHz.  A periodic search was carried out with the pulsar searching software package \texttt{PRESTO}~\citep{2001PhDT.......123R}. The dispersion measure (DM) range that we searched was $0-500\,{\rm pc\,cm}^{-3}$.

\begin{figure}
\plotone{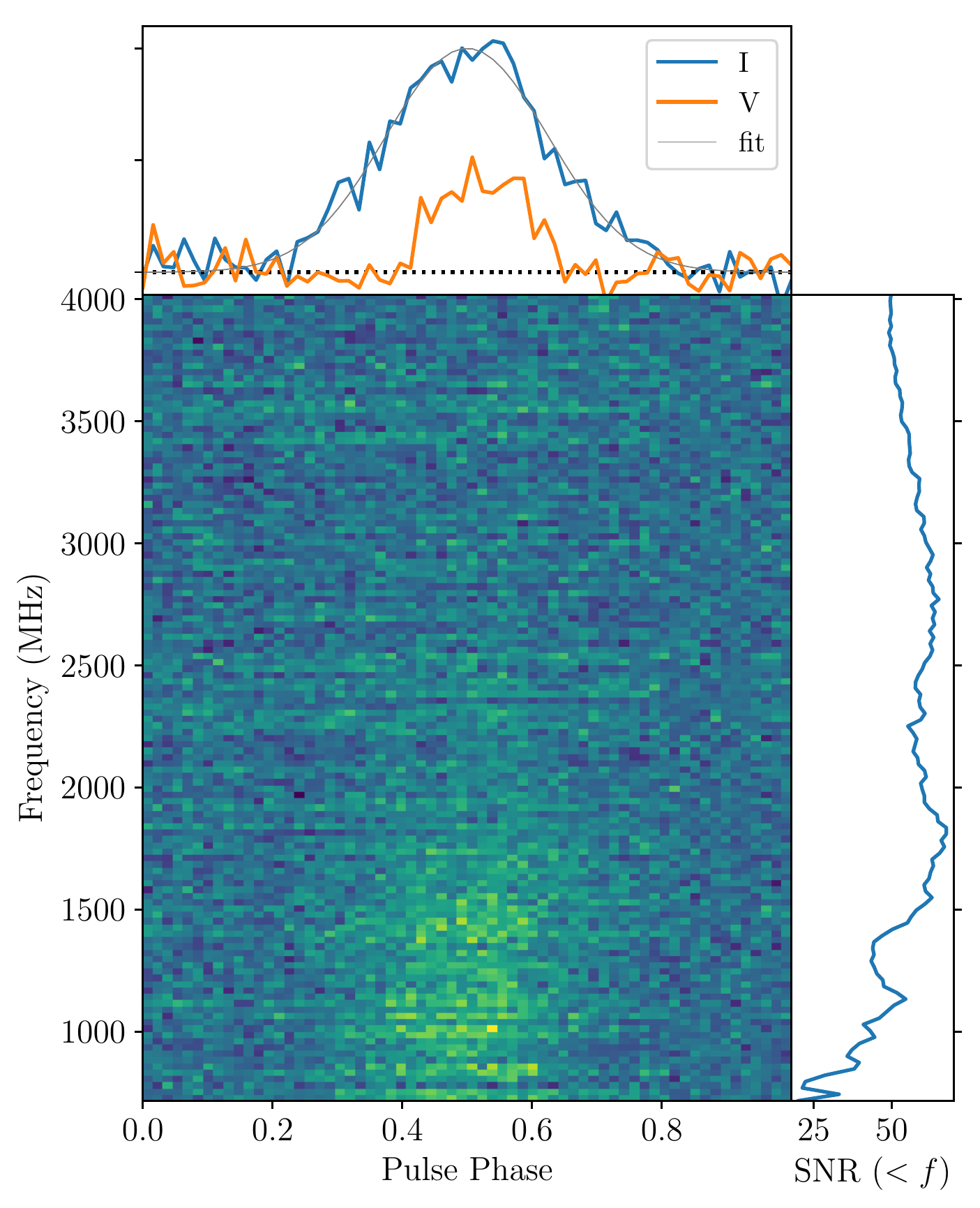}
    \caption{Folded pulse profile of \psr, from the Parkes observation of 2019~May~27.  We show the pulse intensity as a function of frequency over the whole UWL bandpass (lower left), along with the cumulative signal-to-noise ratio as a function of frequency (lower right).  The integrated pulse profile (top) is summed over frequencies $<2\,$GHz, where the signal-to-noise ratio is maximized, along with the best-fit Gaussian.  We show both total intensity (blue) and circular polarization (orange).}
    \label{fig:profile}
\end{figure}

We identified a very strong pulsar candidate with a period of 2.77\,ms
and a dispersion measure ${\rm DM}=228.1\,{\rm pc\,cm}^{-3}$ (see
Figure~\ref{fig:profile}).  This candidate was not in any list of
known pulsars \citep{2016yCat....102034M} or candidates, although it
has since been detected in reprocessed data from the High Time
Resolution-South (HTRU-South) low-latitude survey\footnote{It was
  missed in the original analysis due to a processing error that caused the output to not be inspected, which was later corrected.}\citep[][A.~Cameron, pers.\ comm.]{2015MNRAS.450.2922N}.  Follow-up observations (Table~\ref{tab:obs}) were performed using the coherently dedispersed search mode where data are recorded with 8-bit sampling every 64\,$\mu$s in each of the 1\,MHz wide frequency channels (3328 channels across the whole band). Data were dedispersed at a DM of 228\,${\rm pc\,cm}^{-3}$ with full Stokes information recorded. To measure the differential gains between the signal paths of the two voltage probes, we observed a pulsed noise signal injected into the signal path prior to the first-stage low-noise amplifiers before each observation. A periodic search was carried out for each observation to determine the pulse period at the observing epoch. Data were then folded using the \texttt{DSPSR}~\citep{dspsr} software package. Polarisation calibration was performed as described in \citet{2019ApJ...874L..14D}.

We plot the full UWL bandpass folded on the best-fit pulse period in Figure~\ref{fig:profile}.  It is evident that the pulsar is only visible in the lower half of the band, so we restricted the further analysis to $<2\,$GHz.  We used this bandpass to refine the DM and period measurements.   We find a best-fit DM of $228.27\pm0.02\,{\rm pc\,cm}^{-3}$ and the period measurements in Table~\ref{tab:obs}.

\section{Discussion and Conclusions}
The Parkes observations confirm discovery of a pulsar associated with \askap.  The pulsar is not particularly remarkable: the pulse is broad, with a rectangular equivalent width \citep{2012hpa..book.....L} of 30\%.  A Gaussian model fits well, with full width at half maximum also of 30\% and little apparent change with frequency.  Note that the broad pulse profile may be largely intrinsic: from the scattering relation of \citet{2004ApJ...605..759B}, for ${\rm DM}=228.27\,{\rm pc\,cm}^{-3}$ we expect a scattering timescale of 0.2\,ms at 1.4\,GHz, or less than 10\% of the pulse width.  The pulse is highly polarized, with a peak $V/I$ of about 45\% at the pulse maximum, but the integrated $V/I$ is 21\%, consistent with what we measure from ASKAP.  The polarized fraction is consistent with a constant value across the 700--2000\,MHz band, although there is a hint that it decreases above 2\,GHz.  For a dispersion measure ${\rm DM}=228.27\,{\rm pc\,cm}^{-3}$, we find a distance of 5.0\,kpc with both the NE2001 electron density model \citep{2002astro.ph..7156C} and the YMW16 electron density model \citep{2017ApJ...835...29Y}.  

After rough flux calibration using the system parameters from \citet{UWL}, we find an 800\,MHz flux density of $\approx 0.6\,$mJy, or 0.2\,mJy at 1400\,MHz.  This is considerably lower than what we measure with ASKAP, but seems roughly consistent across the three Parkes observations.  This may indicate variability or unresolved structure in the ASKAP images, but we also caution that commissioning and calibration of both instruments is in progress and that the flux density measurements should be considered provisional.  Diffractive scintillation is an unlikely explanation, as the scintillation bandwidth should be $\ll 1\,$MHz \citep{2002astro.ph..7156C}.  In contrast, the spectral index measured from just the Parkes observations, $-1.66\pm0.07$, is consistent with the ASKAP-only measurement. We will revisit the spectral energy distribution of this source in a later paper.

Note that while this pulsar was detectable in the HTRU observation, the low period and high DM mean that it does rank rather low in ``detectability" \citep{2015MNRAS.450.2185L}, with ${\cal D}=0.03$.  It is not abnormally scattered, nor are there obvious eclipses that would have prevented discovery by a traditional survey.  It does not appear as $\gamma$-ray source in the \textit{Fermi} Large Area Telescope Fourth Source Catalog \citep{2019arXiv190210045T}, which suggests that it is not extremely energetic, but again this is not unusual.  It does mean that it would not have been discovered in the targeted steep-spectrum approach of \citet{2018MNRAS.475..942F}, but there is a wide variety in the  $\gamma$-ray luminosity of pulsars, likely owing to a combination of intrinsic energetics and viewing geometry \citep[e.g.,][]{2010ApJ...716L..85R}.

With multiple Parkes detections, we can measure the period independently in each observation (Table~\ref{tab:obs}).  While within each epoch we see no evidence for a period derivative $\dot P$, non-monotonic changes in the pulsar period of the order of $10^{-6}\,$ms are seen between the Parkes observations.
This compares with a single-epoch uncertainty of at most
$10^{-7}\,$ms.  Note that the period will appear to change secularly
if the position used for barycentering is incorrect
\citep{1972ApJ...173..221M,2008ApJ...675.1468H}.  We can quantify
this, and find that if our position is incorrect by $1\arcsec$ the
period would be incorrect by $10^{-9}\,$ms compared to the
\textit{true} period, and by only $10^{-12}\,$ms from the first
observation to the last.  Therefore this is not the cause of the
period changes.

\begin{figure}
\plotone{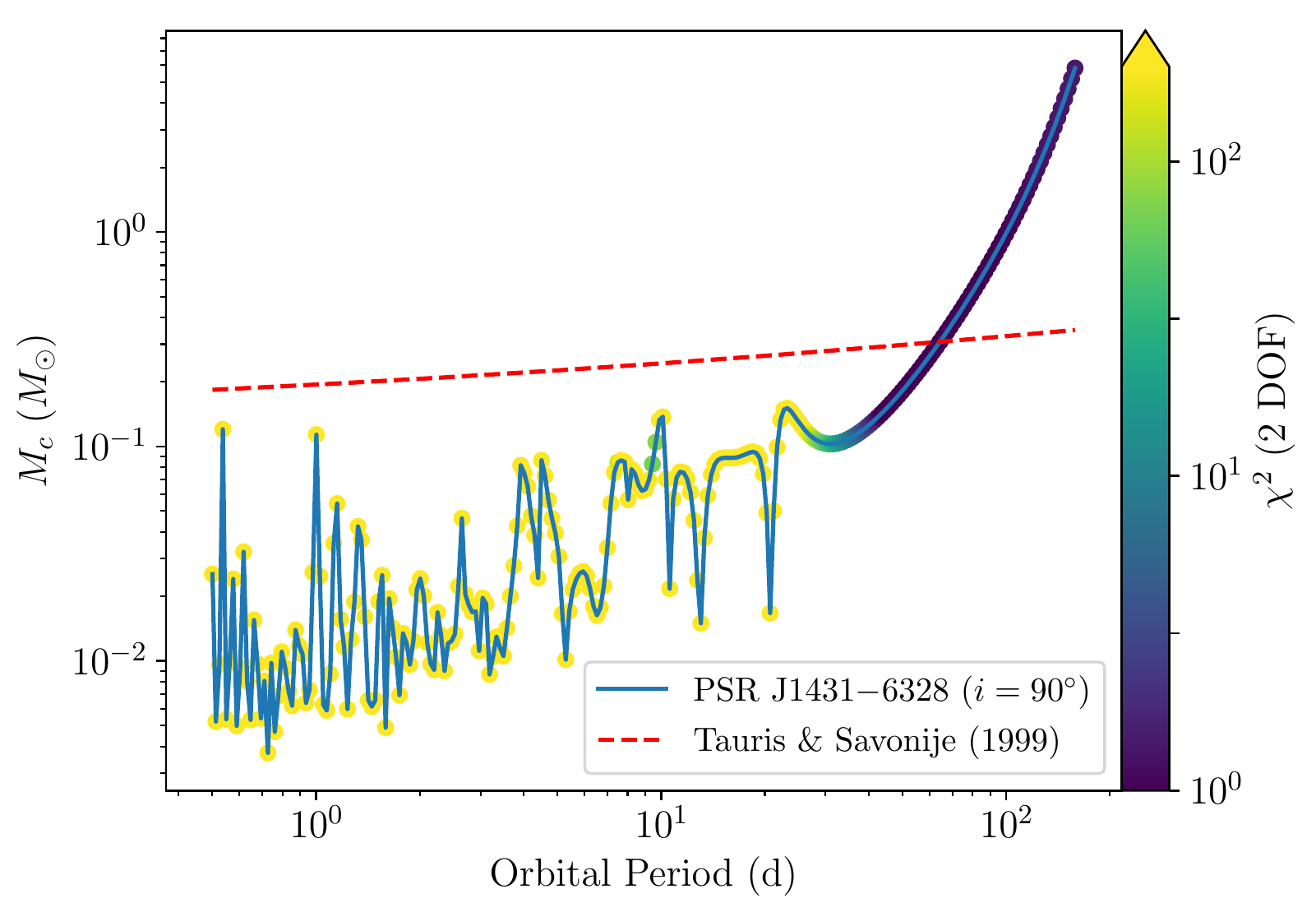}
    \caption{Companion mass $M_c$ constraints as a function of orbital period $P_b$ for \psr, based on the variations in the barycentric pulse period, colored by $\chi^2$ (for 2 degrees of freedom).  This assumes an edge-on, circular orbit and a pulsar mass of $1.4\,M_\odot$; inferred masses for an inclination of $30\degr$ are about a factor of 2 higher.  The constraints for \psr\ are the solid line.  We also show the orbital period-companion mass relation from \citet{1999A&A...350..928T} as the dashed line; this intersects our constraints at $P_b=64.3\,$d and $M_c=0.31\,M_\odot$.  While there are many potential solutions, the long-period solutions have consistently better fits.}
    \label{fig:Mc}
\end{figure}

\begin{figure}
\plotone{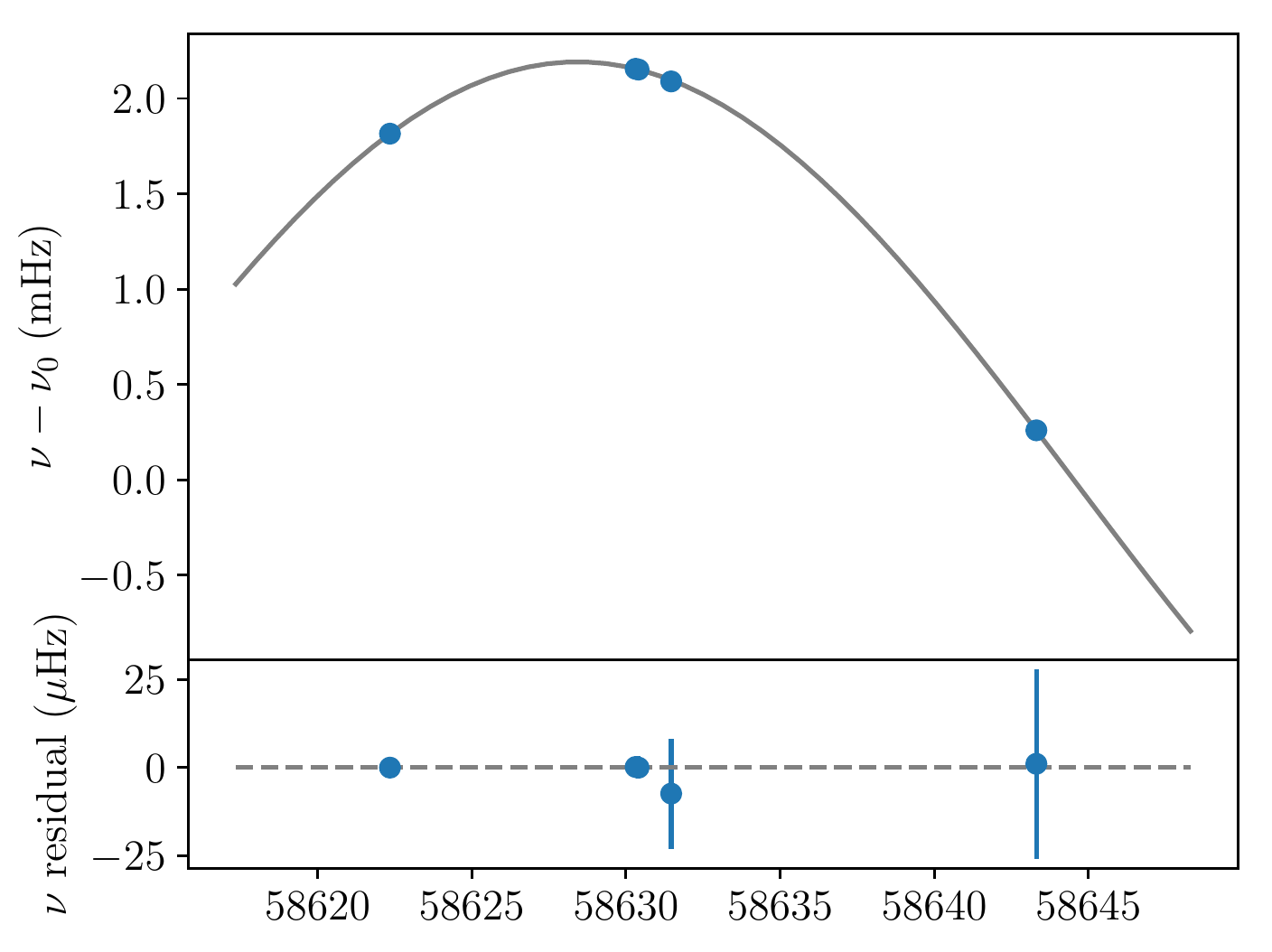}
     \caption{Sample variation of spin frequency with time over the span of observations.  The top panel shows the spin-frequency  values (from Table~\ref{tab:obs}) are indicated by the points, with  $\nu_0$ the best-fit reference frequency.  The bottom panel shows the residuals.  This particular solution is for $P_b=64.3\,$d, which is where the companion mass solutions intersect the predictions of \citet{1999A&A...350..928T}.}
     \label{fig:orbit}
 \end{figure}

If we instead assume that the period changes because of binary motion, we can fit for a range of possible binary orbits. We do not have nearly enough observations for a fully-coherent timing solution, and we cannot even use the incoherent methods like \citet{2008MNRAS.387..273B}.  Instead we iterate over possible binary periods $P_b$ and determine the amplitude of the best-fit circular orbit (since most MSPs are in circular orbits; \citealt{1992RSPTA.341...39P}, although see \citealt{2016ApJ...830...36A} for recent counter-examples) at each $P_b$, fitting for  $P$  in each observation, and marginalizing over the mean spin frequency and orbital phase.  We then convert this to companion masses in Figure~\ref{fig:Mc}.  There are a range of orbits with plausible companion masses, ranging from black widow-like orbits with $P_b<1\,$d and $M_c\sim 10^{-2}\,M_\odot$ \citep{2013IAUS..291..127R} to low-mass helium-core white dwarfs with $P_b=10-100\,$d \citep{1999A&A...350..928T}, although the longer periods have considerably better fits.  Larger masses such as for intermediate-mass binary pulsars \citep{2001ApJ...548L.187C} are excluded, at least for edge-on orbits.
As a specific example solution, our constraint intersects the predictions of \citep{1999A&A...350..928T}  at $P_b=64.3\,$d and $M_c=0.31\,M_\odot$ as shown in Figure~\ref{fig:orbit}.  
Further observations should be able to conclusively determine a timing solution, but for now, the observed period changes are fully consistent with a range of scenarios.  Modest eccentricities like those in \citet{2016ApJ...830...36A} would not likely change our solution dramatically. Note that given the large distance to this source and the modest amount of expected extinction ($A_V\approx 3\,$mag; \citealt{2011ApJ...737..103S}) finding the companion in the optical/infrared seems unlikely, especially as any bright companion (such as a hot, low-mass white dwarf) would have already been identified in the VVV images discussed above.  Nonetheless, once we have a timing position we encourage deeper searches.

There have been a significant increase in new radio surveys recently, with unprecented depth, bandwidth, and frequency coverage.  From low-frequency surveys like the Low-Frequency Array (LOFAR) Two-metre Sky Survey \citep[LoTSS]{2017A&A...598A.104S}, the  GaLactic and Extragalactic All-Sky MWA Survey \citep[GLEAM][]{2015PASA...32...25W}, and the  Giant Metrewave Radio Telescope (GMRT) 150 MHz all-sky radio survey \citep[TGSS]{2017A&A...598A..78I}, to higher-frequency surveys like the Very Large Array (VLA) Sky Survey \citep[VLASS]{2019arXiv190701981L}, the ASKAP Rapid Commissioning Survey (RACS\footnote{\url{https://www.atnf.csiro.au/content/racs}.}; McConnell et al.\ in prep.), and the ASKAP Evolutionary Map of the Universe (EMU; \citealt{2011PASA...28..215N}) project, large amounts of data are now available and the prospects for pulsar discovery are good \citep{2018IAUS..337..328D,2018MNRAS.475..942F,2019arXiv190701981L}.  We wish to ensure that circular polarization is included as part of that search strategy: to our knowledge \psr\ is the first case of a pulsar discovered primarily through circularly-polarized emission.  In contrast to steep-spectrum discoveries only a single survey is needed even for fainter sources, and in contrast to linear polarization foreground depolarization is not an issue and the full band can be used for discoveries.
To assess the potential for finding additional sources like this one in future ASKAP surveys, we consider generic deep (detection threshold $\approx 50\,\mu$Jy) and shallow (detection threshold $\approx 1.5\,$mJy) surveys at 900\,MHz.  {A deep survey would be about 10 times as sensitive as the HTRU high-latitude survey \citep{2010MNRAS.409..619K} in raw flux density}, where the comparison is made with the high-latitude survey despite the low Galactic latitude of \psr\ since we are concerned with future all-sky ASKAP searches.  To quantify this further we created realizations of the millisecond pulsar (MSP) population and the normal (non-recycled) population using \texttt{PsrPopPy2} \citep{2014MNRAS.439.2893B,2015MNRAS.450.2185L}. Of the $ 96$\% of the simulated MSPs and 97\% of simulated normal pulsars with $\delta<+30\degr$ (and hence visible by ASKAP), 66 MSPs and 400 normal pulsars would be detectable by a shallow survey, and 2100 MSPs and 10,000 normal pulsars by a deep survey. Assuming a median circular polarization fraction of $\approx10\%$ \citep{1998MNRAS.300..373H,2018MNRAS.474.4629J}  for both populations, the numbers of projected ASKAP detections via circular polarization alone goes down by a factor of 10: 7 (200) MSPs in a shallow (deep) survey, plus 40 (1000) normal pulsars. Considering that there are currently 262 known MSPs and 1702 normal pulsars \citep{2016yCat....102034M} with $\delta<+30\degr$, it is unlikely that even deep ASKAP searches for circularly-polarized point sources will find large numbers of new pulsars. However, there is a large variation in polarized fraction \citep{2018MNRAS.474.4629J}, and for a 15\% polarized fraction a deep ASKAP survey is as competitive as the HTRU high-latitude survey.
Moreover, MSPs have a larger scale Galactic height compared to normal pulsars \citep{2013MNRAS.434.1387L,2014ApJ...791...67S}, suggesting  that all-sky ASKAP surveys may discover a number of high-$|b|$ MSPs.  Finally, ASKAP could discover MSPs that are hard to find via traditional surveys (highly accelerated such as in \citealt{2018MNRAS.475L..57C}, highly scattered, etc), as discussed above.

\acknowledgments We thank an anonymous referee for comments that
improved this manuscript.  We thank Gregg Hallinan for useful
discussions.  This work was done as part of the ASKAP Variables and
Slow Transients (VAST) collaboration \citep{2013PASA...30....6M}.  DK
was supported by NSF grant AST-1816492.  DK and JS were additionally
supported by the NANOGrav Physics Frontiers Center, which is supported
by the National Science Foundation award 1430284.  TM acknowledges the
support of the Australian Research Council through grant
FT150100099. AZ and DD acknowledge support from an Australian
Government Research Training Program (RTP) Scholarship.  Parts of this
research were supported by the Australian Research Council Centre of
Excellence for All Sky Astrophysics in 3 Dimensions (ASTRO 3D),
through project number CE170100013.  The Australian SKA Pathfinder is
part of the Australia Telescope National Facility which is managed by
CSIRO. Operation of ASKAP is funded by the Australian Government with
support from the National Collaborative Research Infrastructure
Strategy. ASKAP uses the resources of the Pawsey Supercomputing
Centre. Establishment of ASKAP, the Murchison Radio-astronomy
Observatory and the Pawsey Supercomputing Centre are initiatives of
the Australian Government, with support from the Government of Western
Australia and the Science and Industry Endowment Fund. We acknowledge
the Wajarri Yamatji people as the traditional owners of the
Observatory site.  The Parkes radio telescope is part of the Australia
Telescope National Facility which is funded by the Commonwealth of
Australia for operation as a National Facility managed by CSIRO.

\facilities{ASKAP,ATCA,Parkes}

\software{Aegean \citep{2018PASA...35...11H},
  Astropy (\citealt{2013A&A...558A..33A}, \url{www.astropy.org}), CASA \citep{casa}, DSPSR
  (\citealt{dspsr}, \url{http://dspsr.sourceforge.net/index.shtml}), presto (\citealt{2001PhDT.......123R}, \url{http://www.cv.nrao.edu/~sransom/presto/}), PSRCHIVE
  \citep{hotan04},  PsrPopPy2 (\citealt{2014MNRAS.439.2893B}, \url{https://github.com/devanshkv/PsrPopPy2}) PyPulse \citep{2017ascl.soft06011L}, skyfield (\url{http://rhodesmill.org/skyfield/})}.


\end{document}